\documentclass[%
titling,
 twocolumn,
 reprint,
 amsmath,
 amssymb,
 aps,
 nofootinbib
]{revtex4-2}

\usepackage{preamble}

  {\color{red}}%
  {}

\def\pinit{\ensuremath\phi_{\text{e}}}
\def\pfin{\ensuremath\phi_{\text{c}}}

\def\rfin{\ensuremath r_{\text{c}}}

\def\eqprefix{Eq.~}
\def\figprefix{Fig.~}
\def\tabprefix{Table }

\def\SIprefix{\S}
\newcommand\subfigname[1]{.#1}

\DeclareSIUnit \kb{\ensuremath{k_b}}
\DeclareSIUnit \subunit{\text{subunit}}
\DeclareSIUnit\angstrom{\text{Å}}
\DeclareSIUnit\calorie{\text{cal}}

\begin{document}

\author{Noah Toyonaga$^1$, L. Mahadevan$^{1,2}$}
\affiliation{$^1$Department of Physics, Harvard University, Cambridge, MA 02138, USA \\ $^{2}$ Department of Organismic and Evolutionary Biology, and School of Engineering and Applied Sciences, Harvard University, Cambridge, MA 02138, USA}
\title{Structural Dynamics of Contractile Injection Systems}

\begin{abstract}
The dynamics of many macromolecular machines is characterized by chemically-mediated structural changes that achieve large scale functional deployment through local rearrangements of constitutive protein sub-units.
Motivated by recent high resolution structural microscopy of a particular class of such machines, contractile injection systems (CIS), we construct a coarse grained semi-analytical model that recapitulates the geometry and bistable mechanics of CIS in terms of a minimal set of measurable physical parameters.
We use this model to predict the size, shape and speed of a dynamical actuation front that underlies contraction. 
Scaling laws for the velocity and physical extension of the contraction front are consistent with our numerical simulations, and may be generally applicable to related systems.
\end{abstract}

\maketitle


Macromolecular assemblies of biopolymers made of simple repeating subunits are ubiquitious in molecular biology, serving as the building blocks of all biological matter from the cytoskeleton to multi-cellular tissues.
Such assemblies function as reservoirs of information, pathways for transport of matter, and machines that can transform chemical energy into mechanical work~\cite{Mahadevan2000-zb}. 
In this last role, conformational changes at the level of a constitutive subunit can be used to both regulate (slow) assembly and (fast) deployment in which bending and twisting at small scales is amplified geometrically to shrinkage and extension of the global structure.
Prominent examples of these biological engines include polymorphic bacterial flagella, which transition between multiple shapes reversibly~\cite{Kamiya1979-mw, Calladine2013-ey,srigiriraju2006model}, viral capsids, where self assembly is regulate by a conformational change in the monomer catalyzed by the appropriate growth site \cite{Caspar1980-ob}, acrosomal reactions in marine invertebrates such as the horse-shoe crab wherein an initially twisted actin bundle straightens in the presence of $\text{Ca}^{2+}$ \cite{Mahadevan2011-wj}, and polymorphic transitions in a variety of cytoskeletal assemblies \cite{kan2007shape,ryan2023bend} etc. 

Recently, there has been much interest in a particular class of macromolecular machines known as Contractile Injection Systems which includes Myoviridae bacteriophage tails (T4), R-type Pyocins, Type VI Secretion Systems (T6SS), and phage-like protein-translocation-structures (PLTS) \cite{Clemens2015-nk,Ge2015-xx,Kudryashev2015-pj,Kostyuchenko2005-of,Desfosses2019-mn, Wang2017-mb,Guerrero-Ferreira2019-on}.  These assemblies function as spring-loaded needle mechanisms that can be deployed by their host organism to puncture  membranes, and serve as the physical pathway for infection.

All CIS consist of three components---a bistable sheath made of interlaced protein strands, a hollow tube, and a baseplate. 
Prior to deployment the hollow tube is contained within the sheath and the combined assembly is capped by the baseplate. 
{\em In-vivo} the baseplate mediates interactions with target substrates (e.g. the membrane of a target cell): upon binding to the target, it dramatically changes conformation \cite{Coombs1994-xa,Kostyuchenko2005-of}, thereby initiating the global contraction of the sheath.  
In turn, the sheath pushes the tube out through a hole in the baseplate with sufficient force to puncture the target membrane, as shown schematically in Fig. 1(a-b). 
In this work, we are concerned with the dynamics of the sheath contraction process; that is, how a local reconfiguration of the sheath proteins (initiated by baseplate reconfiguration on adhesion with a host) propagates through the sheath system. 

High resolution structural microscopy \cite{Wang2017-mb, Desfosses2019-mn} reveals the nature of chemical interactions between the strands which constitute the sheath, and which are critical for both the assembly and controlled deployment of the CIS. 
These structural studies indicate that variations in the non-covalent bonding between strands (on the order of 10 kcal/mol per strand subunit) can serve to both stabilize the CIS in a high energy metastable state and also provides the energy to drive contraction \cite{Ge2015-xx, Aksyuk2009-wp}. Thus, the reconfiguration of strands during contraction gives rise to a surplus free energy due to changes in surface bonding which in turn drives deployment via a self-propagating process\cite{Taylor2018-fn}.
These computational results are also consistent with earlier measurements of the free energy released during deployment in in-vitro experiments \cite{Arisaka1981-uq}.

It is useful to note that contraction of the sheath can also be triggered {\em in-vitro} by treatment of CIS sheaths with urea \cite{Arisaka1981-uq, Leiman2004-fm}, elevated temperature \cite{Arisaka1981-uq}, acid \cite{Coombs1994-xa}, and cationic degergents \cite{To1969-es}, \textcolor{black}{all of which are hypothesized to disturb the aforementioned strand-strand bonding \cite{Aksyuk2009-wp, Leiman2012-ps}. }

The Angstrom-level atomic resolution structures of CIS in the extended and contracted states naturally raises the question about the dynamical principles underlying their deployment. Recent works have sought to understand these dynamics using large-scale molecular computational methods \cite{Fraser2021-zu} but have been hampered by the large physical scale of the structure (with molecular weights of several megadaltons), and the concomitant global nature of the deformation.   Coarse-grained dynamical models \cite{Maghsoodi2017-lt} represent the CIS sheath as a set of elastic strands interacting with an ambient fluid but do not account for the metastability of the CIS and therefore only describe the over-damped relaxation of the system \cite{Maghsoodi2019-ym}.
(For a survey of related prior work see SI)
Thus, even basic questions such as whether the contraction proceeds uniformly everywhere or propagates like a wave through the structure remain unanswered \cite{Maghsoodi2019-ym, Maghsoodi2017-lt}. 

Here we develop a coarse-grained dynamical model for the extrusion of CIS based on structural observations by 1)  parameterizing the geometry of the CIS using a Chebyshev net (discussed in detail below), which allows us to project the dynamics of the full three dimensional structure onto a scalar order parameter which measures the local deformation, and
2) accounting for the (surface-bonding induced) meta-stability of the CIS using a bistable potential whose free parameters are fixed by experimental measurements.
The resulting framework provides scaling predictions for the speed and nature of the dynamics of the CIS contraction process while elucidating general principles for the dynamics of similar macromolecular assemblies.

\begin{figure*}
	\centering
    \includegraphics[width=1\linewidth]{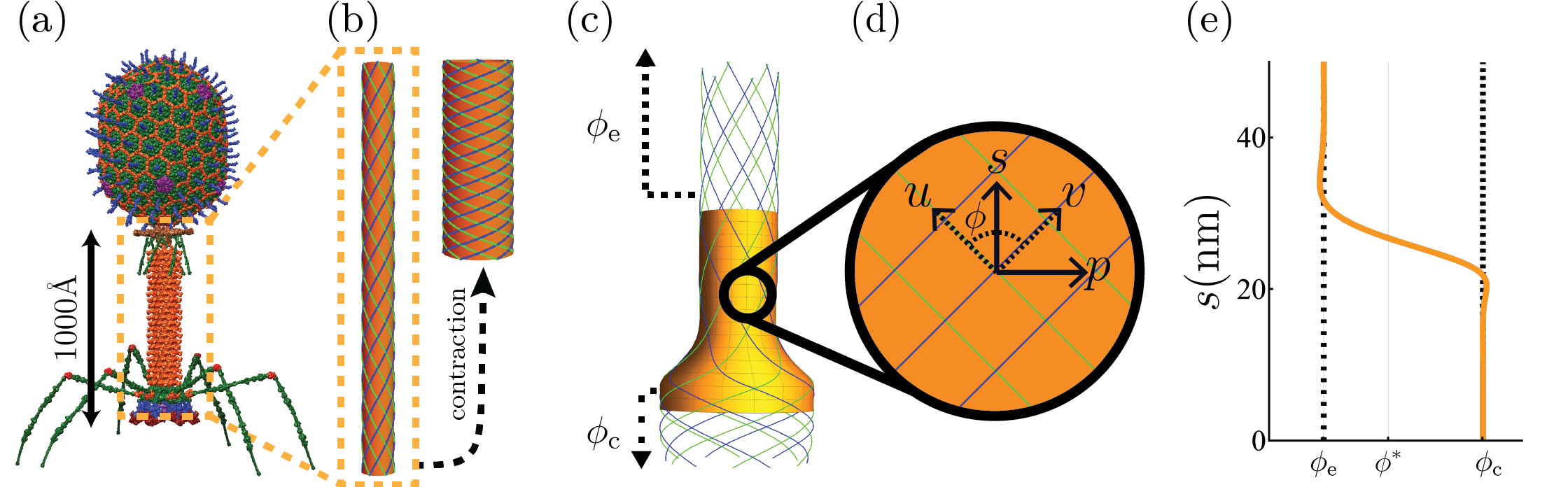}
  \caption{\textbf{CIS atomic structure and coarse-grained model}:
  (a) The tail of a T4 bacteriophage is an example of a contractile injection system which is used to infect target cells.
  (b) The protein sub-units which constitute a CIS are interlinked to form a lattice. 
  This motivates us to model the geometry of an arbitrary CIS sheaths as an axisymmetric surface $\Phi(u,v)$ whose coordinate curves (\{$\partial_u\Phi$, $\partial_v\Phi$\}, highlighted in green and blue) are parallel to the protein strands of the modeled structure. 
  The opening angle between strands is the dynamical order parameter in our model which we denote by $\phi(s, t)$. 
  The extended and contracted configurations of the proposed continuum model correspond to constant values of the order parameter $\phi$: $\phi(s,t) = \pinit$ and $\phi(s,t) = \pfin$, respectively.
  (c) A representative contraction front represents a transitional region between two domains of stable states, $\pinit$ above and $\pfin$ below. (d) The 3D geometry of the sheath depends on the angle $\phi$ between the strand axes $\{{u},{v}\}$.
  We work in a coordinate system $\{ {s}, {p}\}$ defined relative to the strand axes  
  as shown in the zoom-in and described in the SI.
  (e) Plot of the order parameter $\phi(s,t=t')$ corresponding to the CIS geometry shown in (c). 
  The dotted vertical lines at $\pinit$, $\pfin$ correspond to the initial and final states of the CIS (shown in (b)). 
  Note the horizontal axis here is in terms of the material coordinate $s$ rather than $z$, the vertical coordinate of the embedded geometry (shown in (c)).
}
\label{fig:structure_and_model}
\end{figure*}

\label{sec:CIS_Geometry}

The conserved sheath structure of the CIS consists of identical protein sub-units which are linked in a mesh of helical strands which are connected to one another, akin to the warp and weft fibers in a woven fabric \cite{Aksyuk2009-wp, Ge2015-xx, Wang2017-mb, Desfosses2019-mn, Caspar1980-ob}. 
This motivates our coarse-grained model for the CIS sheath geometry as a two dimensional surface $\Phi(u,v)$ whose local tangent vectors ($\partial_{u}\Phi, \partial_{v}\Phi$) are everywhere parallel to the strands of the corresponding CIS structure. Consistent with structural information \cite{Ge2015-xx}, we assume that the relative distance between sub-units are fixed by inextensible linkages, but remain free to rotate about one another, allowing us to parameterize the kinematics of deformation by the angle between the coordinate vectors in the embedded configuration.
This allows us to describe the local conformation of the CIS in terms of a single quantitiy, the shearing angle $\phi \equiv \phi(s, p, t)$ between strands as a function of the axial material coordinate $s$ and an azimuthal material coordinate $p$  and time $t$ (see \figprefix{}\ref{fig:structure_and_model}\subfigname{d}) and serves as a  scalar order parameter for the transition between the contracted and extended states.

If we assume that the surface is axisymmetric then the angle $\phi=\phi(s,t)$  is independent of $p$, and at any instant of time $t$ we can construct the following explicit embedding for $\Phi$ parameterized by $\phi$ 
(see SI  for details):

\begin{subequations}
\label{eq:phi_general}
\begin{equation}
    \Phi(s, p,t) = \begin{bmatrix}
           r(s,t)\cos(p/r_0)) \\
           r(s,t)\sin(p/r_0)) \\
           z(s)
         \end{bmatrix}
\end{equation}
\begin{equation}\label{eq:phi_general_b}
    r(s,t) = r_0\sin(\phi(s,t)/2)
\end{equation}
\begin{equation}
     z(s,t) = \int_0^s \sqrt{\cos ^2\left(\frac{\phi \left(s',t\right)}{2}\right)-\left(\frac{dr}{ds}\big|_{s'}\right)^2} \, ds'
\end{equation}
\end{subequations}

Here $r_0$ in \eqprefix{}\ref{eq:phi_general_b} is fixed by the radius of the CIS. 
When the CIS is its initial and final cylindrical states the shearing order parameter $\phi(s,t)$ takes the constant value $\phi(s,t)=\pinit, \pfin$ respectively and the corresponding surface $\Phi$ is a cylinder of radii $r=r_0\sin(\pinit/2), r=r_0\sin(\pfin/2) $ with helical coordinate curves as shown in \figprefix{}\ref{fig:structure_and_model}\subfigname{b}.
Structural data allows us to explicitly assign the initial and final geometric states in our model: e.g. given the radius $\rfin$ and shearing angle $\pfin$ of a CIS in the contracted state, it follows that $r_0\equiv \frac{\rfin}{\sin(\pfin/2)}$  \cite{Desfosses2019-mn,Wang2017-mb}.

To move towards the energetics and dynamics of the CIS parameterized by the geometry of the shearing order parameter $\phi(s,t)$,
we must account for both 1) mechanical stresses in the strands induced in the sub-unit strands as they are deformed from the initial to final configurations and 2) changes in the chemical potential due to the formation and breakage of inter-molecular bonds between sub-units as they change conformation. 
We model the strands of the sheath as anisotropic elastic filaments that can bend only in one direction, with a local energy density $EI\kappa(\phi)^2/2$ where $E$ is the Young's modulus, $I$ is the second moment of area, and $\kappa(\phi)$ is the filament curvature. Modeling the chemical energy associated with the changes in the shear angle via potential $U(\phi)$ (discussed below), 
leads to an expression for the total energy of the CIS given by: 
\begin{equation}\label{eq:free_energy}
	F[\phi] = \int ds \bigg[\underbrace{EI\kappa(\phi)^2}_{\text{Bending}} + \underbrace{U(\phi)}_{\text{Shear}} \bigg]
\end{equation}
To characterize the terms in \eqprefix{}\ref{eq:free_energy}), we first note that the strands of the CIS lie along the material curves $u$ and $v$ of the map $\Phi(u,v)$, and primarily bend out of plane but are stiff in-plane (i.e. mechanically they are ribbon-like). Then, the curvature of each rod is equivalent to the normal curvature of the surface $\Phi$ parallel to these material curves given by (see \SIprefix\ref{SI:curvature} for details). 
%
If the shear order parameter $\phi(s,t)$ is slowly varying (i.e. $ \left(\frac{d\phi}{ds}\right)^2< \left(\frac{1}{2r_0}\right)^2$), we can drop higher order terms in $\phi'\equiv \partial \phi(s,t)/\partial s$, leading to a simplified expression for the curvature
\begin{equation}\label{eq:simplified_normal_curvature}
    \kappa (\phi) = \underbrace{-\frac{\sin \left(\frac{\phi (s,t)}{2}\right)}{r_0}}_{\text{helix curvature}} + \underbrace{\frac{1}{2} r_0 \phi ''(s,t) \cos \left(\frac{\phi (s,t)}{2}\right)}_\text{gradient contribution}
\end{equation}
which is the form we use \footnote{As previously discussed the initial and final stable configurations of the CIS sheath corresponds to helices (characterized by the opening angle $\phi_i$, $\phi_f$); thus, the gradient contribution will only be active in the region of the shear wave front. Given the radius of the CIS $\approx\qty{100}{\angstrom}$, our assumption of a slowly-varying shear angle $\phi(s,t)$ corresponds to a contraction wave of length $\approx\qty{50}{\angstrom}$, or $1/20$ the length of the structure ($\approx\qty{1000}{\angstrom}$), and is consistent with experimental observations of the assembly in its contracted and expanded states\cite{Ge2020-kx}.}, and which can be interpreted as a sum of two terms,  associated with the curvature of a helix of constant radius and a higher order gradient correction.

\begin{figure*}[ht]
	\centering
  \includegraphics[width=.9\linewidth]{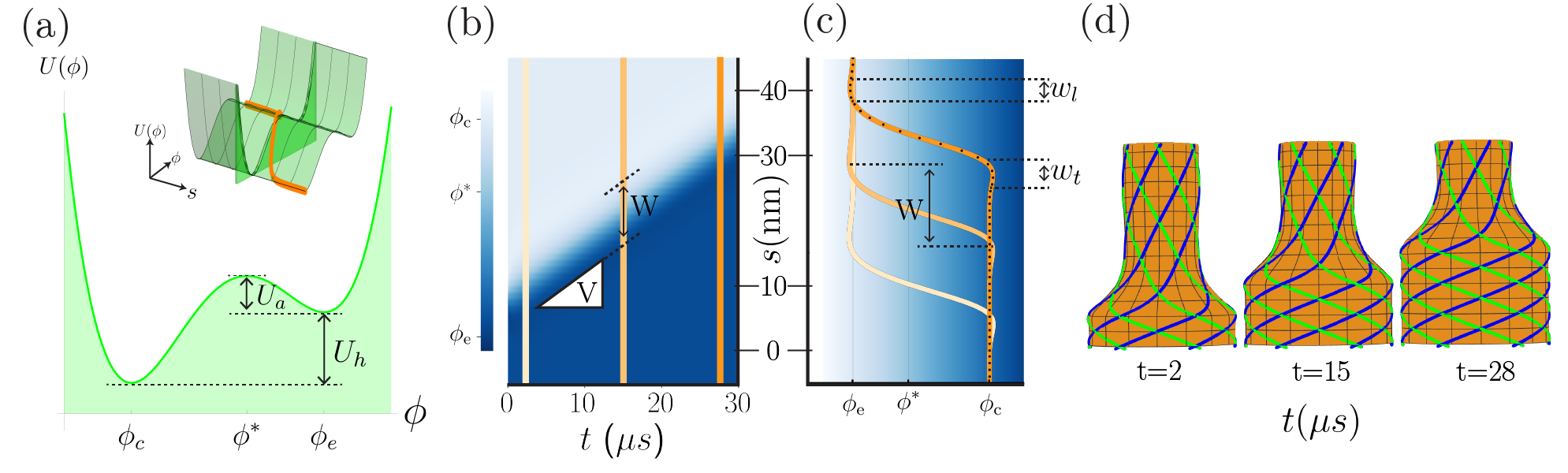}
  \caption{\textbf{Propagating front of conformational change}: (a) The quartic potential $U(\phi)$ (defined by \eqprefix{}\ref{eq:potential}) drives the evolution of the shearing order parameter $\phi(s,t)$. 
  The 2D plot shows the geometric ($\pinit$, $\pfin$) and energetic quantities ($U_h$, $U_a$) we use to fix the free parameters of $U$ (\eqprefix{}\ref{eq:potential}) based on experimental measurements (see footnote 2 for description).
  The inset 3D plot shows a static snapshot of the shearing order parameter $\phi(t=t^*, s)$ (drawn in orange) superimposed on the potential, which is constant along the axial direction, $s$.
  (b) The shearing order parameter $\phi(s,t)$ from the simulation described in the main text and indicated by a star in \figprefix\ref{fig:scaling_relations} plotted on the $t, s$ plane. 
  Color corresponds to the value of $\phi$ in radians. 
  The diagonal band corresponds to the front moving at a constant speed speed $\text{V}$ and with a characteristic width $W$.
  (c) Plot of $\phi(t=t^*, s)$ taken at time steps  $t^*\in\{\qty{2}{\micro\second}, \qty{15}{\micro\second}, \qty{28}{\micro\second}\}$ (corresponding to vertical highlights in (b)).
  Again, we observe the geometry of the front is constant  during propagation (i.e. the different lines are identical up to translation).
  We characterize the shape of the front by measuring the width $W$, and leading and trailing ridge widths ($w_l$, $w_t$).  
	(d) The three dimensional geometry of the front (as a segment of the CIS) corresponding to $\phi(s,t)$ at $t=2, 15, 28 \unit{us}$ (these snapshots correspond to the same solutions highlighted in (b) and (c)).
}
\label{fig:sample_solution}
\end{figure*}

We now turn to the chemical free energy in \eqprefix{}\ref{eq:free_energy} associated with the changes in the non-covalent bonding between strands in the CIS sheath that gives rise to a bistability of the structure. We model this with a simple double well potential with units of energy density (\unit{\joule\per\meter})  illustrated in \figprefix{}\ref{fig:sample_solution}\subfigname{a}, and given by

\begin{equation}
	\label{eq:potential}
	\begin{split}
		U(\phi) =& b\left(a\left( c\phi - \phi_0\right) - \frac{1}{2}\left( c\phi - \phi_0\right)^2 - 
		\right . \\
			 & \left. \frac{a}{3}\left( c\phi - \phi_0\right)^3 
			 + \frac{1}{4}\left( c\phi - \phi_0\right)^4\right)
	\end{split}
\end{equation}

We note that \eqprefix{}\ref{eq:potential} is the simplest form which captures the essential chemical-mechanics  of CIS
and is \textit{uniquely determined} from structural and calorimetric experiments of CIS \cite{Ge2015-xx, Ge2020-kx, Fraser2021-zu}  with no free parameters 
\footnote{
A brief summary of our procedure for fitting the parameters of $U(\phi)$ to experimental data is provided here, while a longer discussion is in \SIprefix\ref{SI:model_free_parameters}. 
We note that the parameters of \eqprefix{}\ref{eq:potential} ($a, b, c, \phi_0$) set the location of the two minima of $U$ ($\pinit$, $\pfin$), the potential difference between the minima ($U_h$), and the height of the potential barrier dividing them ($U_a$) as shown in \figprefix{}\ref{fig:sample_solution}\subfigname{a}. We assign the minima of $U$ to $\pinit$ and $\pfin$ so that the dynamics of $\phi(s,t)$ correspond to flow of the system from the extended, high energy metastable state ($\pinit$) to the contracted, low energy state ($\pfin$) (see \figprefix{}\ref{fig:sample_solution}\subfigname{a}, inset).
Similarly, we assign $U_h$ to be the enthalpy density and $U_a$ corresponds to the free energy density necessary to initiate contraction, both of which can be measured from calorimetry or estimated from structural models.
Here, $U_a$ and $U_h$ have units of \unit{\joule\per\meter}, since the total energy of the system is integrated over the length of the CIS (see \eqprefix{}\ref{eq:free_energy}).
For the simulations presented in this work we take $\pinit=\qty{1}{\radian}$, $\pfin=\qty{2.4}{\radian}$ and vary $\qty{1e-11}{\joule\per\meter}<U_h<\qty{4e-11}{\joule\per\meter}$ and $\qty{1e-11}{\joule\per\meter}<U_a<\qty{4e-11}{\joule\per\meter}$, corresponding to the geometry and energetics of the R2-type Pyocin \cite{Ge2015-xx, Ge2020-kx, Fraser2021-zu}.
}.

To derive equations of motion for the CIS from the free energy given in \eqprefix{}\ref{eq:free_energy}, we note that the small-scale nature of the CIS (10s of \unit{\nano\meter}) suggests that its dynamical evolution follows from a balance between the internally stored energy and viscous dissipation. Balancing local torque densities implies that $ \eta \frac{\partial \phi(s,t)}{\partial t} = -\frac{\delta F}{\delta \phi}$, where $\eta$ is the effective friction factor with dimensions of \unit{\newton\second}, leading to the following equation of motion for the evolution of the shear order parameter $\phi(s,t)$ (details in SI):

\begin{equation}\label{eq:full_pde}
	\begin{split}
		\eta \frac{\partial \phi}{\partial t} =& -EI\bigg(\frac{\sin(\phi)}{2r_0^2} + \left(\frac{1}{2}\sin\left(\phi\right)\left(\phi'\right)^2 - \cos\left(\phi\right)\phi''\right)+ \\
			   & \frac{r_0^2}{8}\left(-2\cos(\phi)\left(\phi'\right)^2\phi''
		   -3\sin(\phi)\left(\phi''\right)^2 \right. \\
			   & \left. -4\sin(\phi)\phi'\phi''' + 2\cos(\phi)\phi''''+2\phi'''' \right)\bigg) - f(\phi)
\end{split}
\end{equation}
where $f(\phi)\equiv\frac{\partial U}{\partial \phi}$.


We observe that our derivation of \eqprefix{}\ref{eq:full_pde}  follows from the assumption that local shear rates $\phi_{t}$ are more important than their spatial derivatives $\propto\frac{\partial^2\phi}{\partial t \partial s}$.
The latter choice would not give rise to a constant velocity front solution (physically the CIS would, in general, fail to fully actuate).
We also note out that the formulation here is equivalent to that used to explain the dynamics of other macromolecular assemblies such as the acrosome reaction \cite{Mahadevan2011-wj}, where the dynamical equations take the form $ \frac{\partial^2\phi}{\partial t \partial s} = -\frac{\partial }{\partial s}(\frac{\delta F}{\delta \phi})$, but for a  different form of the free energy (as a function of the twist strain), which on integration yields the same form as here.

To follow the dynamics of the contraction process, we search for solutions of the high-order nonlinear PDE (\eqprefix{}\ref{eq:full_pde}) that connect the stationary solutions (i.e. solutions in which $\phi$ transitions from the extended state $\phi(-\infty,t)-\pinit=\phi'(\infty,t)=0$ to the contracted state $\phi(\infty,t)-\pfin=\phi'(\infty,t)=0$). 
We use a Chebyshev spectral method  \cite{Trefethen2000-wg, burns2020} to solve (\eqprefix{}\ref{eq:full_pde}), with an initial condition corresponding to a step function convolved with a Gaussian with width $\sigma$ of the same order as $r_b$. 
This initial condition physically corresponds to a state where one end of the CIS has been forced into its collapsed, low energy state; in vivo, this maps to the initial reconfiguration of the lower sheath sub-units by the base plate upon binding to a target substrate \cite{Ge2020-kx, Guerrero-Ferreira2019-on}.


In \figprefix\ref{fig:sample_solution}\subfigname{b}, we show the results of a representative simulation and observe that the order parameter $\phi(s,t)$ settles into a steadily propagating front with a constant velocity $\text{V}$ and width $\text{W}$ (defined as the distance between the minimum and maximum values of $\phi$ at a fixed time). We also observe that the front is preceded/succeeded by asymmetric oscillatory ridges when $\phi<\pinit$ and $\phi>\pfin$, as expected from the asymptotic behavior of these high-order PDEs. We define the trailing width $w_l$ as the full width half maximum (FWHM) between $\min{\phi}$ and $\pfin$ (and analogously define $w_t$ as the FWHM between $\max\phi$ and $\pinit$) as illustrated in \figprefix\ref{fig:sample_solution}\subfigname{c}.

\begin{figure}
	\includegraphics[width=.8\linewidth]{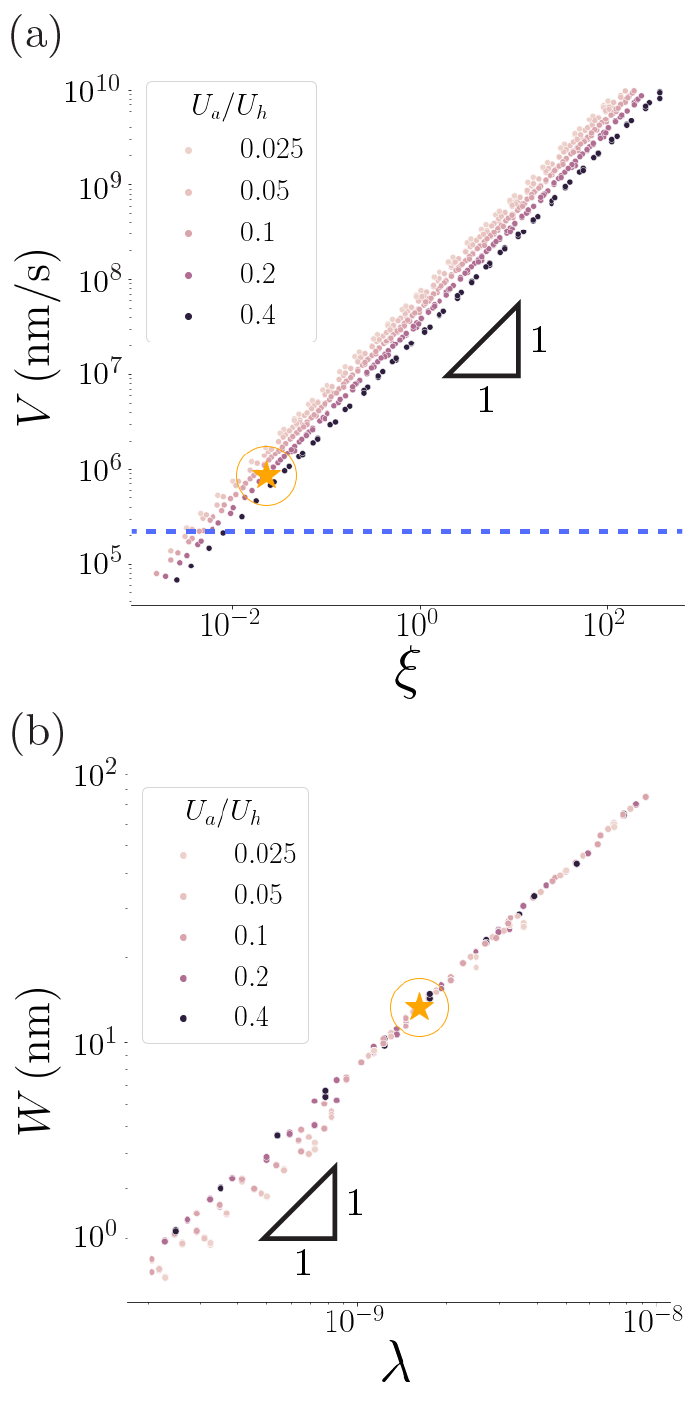}
	\caption{\textbf{Scaling laws for speed and width of front}: We simulate 946 CIS systems using parameters given in the SI
     by numerically solving \eqprefix{}\ref{eq:full_pde}.
     From each simulation we extract the (a) velocity ($V$) and (b) width ($W$) of the propagating front (see \figprefix{}\ref{fig:sample_solution}) and plot each against $\xi$ and $\lambda$ (see \eqprefix{}\ref{eq:scaling_width}, \ref{eq:scaling_velocity}), observing they collapse onto a line consistent with our physical derivation of these scaling parameters.
	The orange star on each plot indicates the simulation illustrated in \figprefix\ref{fig:sample_solution}.
	The blue line in (a) indicates the lower bound (\qty{2.0e5}{\nano\meter\per\second}) for the velocity consistent with experimental data (see SI).}
\label{fig:scaling_relations}
\end{figure}

To understand the nature of the propagating front solution arising from \eqprefix{}\ref{eq:full_pde} we consider a balance between the viscous drag ($\eta \partial \phi/\partial t$), chemical forcing ($\sim \tilde f \equiv f(\phi\approx \pi/2))$, and generalized mechanical forces dominated by the highest order term ($\sim {EI}{r_0^2}\phi''''$) at the front \footnote{This is equivalent to having substantial curvature gradients  (see Eq. (3)) in the neighborhood of the front.}. Balancing the second and third, i.e. the chemical forcing and mechanical strain yields a characteristic width over which $\phi$ varies at the shearing front:

\begin{equation}
\label{eq:scaling_width}
		\lambda \sim \left(\frac{EIr_0^2}{\tilde{f}}\right)^{1/4}
\end{equation}


Balancing the first and second leads to a time scale $\tau \sim \eta/\tilde{f}$, which combined with the previous length scale (or equivalently stating that all three terms must balance each other) yields a scaling law for the front velocity: 

\begin{equation}\label{eq:scaling_velocity}
		\xi  \sim \frac{\lambda}{\tau} \sim \frac{\left(\tilde{f}^3EIr_0^2\right)^{1/4}}{\eta}
\end{equation}







In \figprefix\ref{fig:scaling_relations} we show that numerical simulations of \eqprefix{}\ref{eq:full_pde} yields the front width and velocity that follow the above scaling laws, consistent with our assumption that the geometry and dynamics of the CIS traveling front are controlled by a balance between viscous friction, and chemo-mechanical forces.

\label{sec:comparison_to_experiment}

Using experimentally available parameter values 
(see  \tabprefix{}\ref{tab:physical_parameters}) 
for the geometry of the CIS strands 
($r_0 \approx \qty{100}{\angstrom}$, $r_b \approx \qty{10}{\angstrom}$), elastic stiffness of biological filaments ($E \approx \qty{1E9}
{\Pa}$), activation and enthalpy densities ($U_a \approx \qty{2E-12}{\joule\per\meter}$, $U_H \approx \qty{2E-11}
{\joule\per\meter}$), and a dynamic shearing friction coefficient $\eta \approx \qty{1E-20}{N.s}$ (see SI for a discussion for how this was estimated), we recover a front width of $l \approx \qty{60}{\angstrom}$ and a velocity of $v \approx \qty{1e-3}{\meter\per\second}$. We note that this front width is consistent with the assumption of a slowly varying shear used in deriving the simplified bending energy density, while the velocity is larger than the lower bound ($v_\text{min} = \qty{2E-4}{\meter\per\second}$) estimated from experiments (see \S S.IV for details). Importantly, we note that the friction coefficient used in our simulations corresponds to an effective viscosity of the surrounding fluid of $\eta\approx \qty{1e-2}{\pascal\second}$, which is well within the range of effective viscosities observed in nanoscale systems. 

\label{conclusion}

Our model has focused on the physico-chemical mechanisms associated with the CIS. However, the interplay between a chemically-induced bistability of sub-unit conformation, the mechanical stiffness of protein strands, and viscous stresses from the ambient surroundings are features generic to a larger class of macromolecular machines. It is plausible that in many such machines, the formation of a critical nucleus (the binding of baseplate to target in the case of CIS) will be followed by a front of geometric rearrangement which propagates at constant velocity. Although the evolutionary history and biological function of such structures may differ, their deployment is likely to be controlled by scaling relations similar to those of $\lambda$, $\xi$ (\eqprefix{}\ref{eq:scaling_width}, \ref{eq:scaling_velocity}), a perhaps general principle worth pondering further.

\bibliography{references}

\begin{thebibliography}{10}

\bibitem{Mahadevan2000-zb}
L~Mahadevan and P~Matsudaira.
\newblock Motility powered by supramolecular springs and ratchets.
\newblock {\em Science}, 288(5463):95--100, April 2000.

\bibitem{Kamiya1979-mw}
R~Kamiya, S~Asakura, K~Wakabayashi, and K~Namba.
\newblock Transition of bacterial flagella from helical to straight forms with
  different subunit arrangements.
\newblock {\em J. Mol. Biol.}, 131(4):725--742, July 1979.

\bibitem{Calladine2013-ey}
C~R Calladine, B~F Luisi, and J~V Pratap.
\newblock A ``mechanistic'' explanation of the multiple helical forms adopted
  by bacterial flagellar filaments.
\newblock {\em J. Mol. Biol.}, 425(5):914--928, March 2013.

\bibitem{srigiriraju2006model}
Srikanth~V Srigiriraju and Thomas~R Powers.
\newblock Model for polymorphic transitions in bacterial flagella.
\newblock {\em Physical Review E}, 73(1):011902, 2006.

\bibitem{Caspar1980-ob}
D~L Caspar.
\newblock Movement and self-control in protein assemblies. quasi-equivalence
  revisited.
\newblock {\em Biophys. J.}, 32(1):103--138, October 1980.

\bibitem{Mahadevan2011-wj}
L~Mahadevan, C~S Riera, and Jennifer~H Shin.
\newblock Structural dynamics of an actin spring.
\newblock {\em Biophys. J.}, 100(4):839--844, February 2011.

\bibitem{kan2007shape}
Wanxi Kan and Charles~W Wolgemuth.
\newblock The shape and dynamics of the leptospiraceae.
\newblock {\em Biophysical journal}, 93(1):54--61, 2007.

\bibitem{ryan2023bend}
Paul~M Ryan, Joshua~W Shaevitz, and Charles~W Wolgemuth.
\newblock Bend or twist? what plectonemes reveal about the mysterious motility
  of spiroplasma.
\newblock {\em Physical review letters}, 131(17):178401, 2023.

\bibitem{Clemens2015-nk}
D~L Clemens, P~Ge, B-Y Lee, M~A Horwitz, and Z~H Zhou.
\newblock Atomic structure of {T6SS} reveals interlaced array essential to
  function.
\newblock {\em Cell}, 160(5):940--951, February 2015.

\bibitem{Ge2015-xx}
P~Ge, D~Scholl, P~G Leiman, X~Yu, JF~Miller, and Z~H Zhou.
\newblock Atomic structures of a bactericidal contractile nanotube in its pre-
  and postcontraction states.
\newblock {\em Nat. Struct. Mol. Biol.}, 22(5):377--382, May 2015.

\bibitem{Kudryashev2015-pj}
M~Kudryashev, R~Wang, M~Brackmann, S~Scherer, T~Maier, D~Baker, F~DiMaio,
  H~Stahlberg, E~H Egelman, and M~Basler.
\newblock Structure of the type {VI} secretion system contractile sheath.
\newblock {\em Cell}, 160(5):952--962, February 2015.

\bibitem{Kostyuchenko2005-of}
V~A Kostyuchenko, P~R Chipman, P~G Leiman, F~Arisaka, V~V Mesyanzhinov, and M~G
  Rossmann.
\newblock The tail structure of bacteriophage {T4} and its mechanism of
  contraction.
\newblock {\em Nat. Struct. Mol. Biol.}, 12(9):810--813, August 2005.

\bibitem{Desfosses2019-mn}
A~Desfosses, H~Venugopal, T~Joshi, J~Felix, M~Jessop, H~Jeong, J~Hyun, J~B
  Heymann, M~R~H Hurst, I~Gutsche, and A~K Mitra.
\newblock Atomic structures of an entire contractile injection system in both
  the extended and contracted states.
\newblock {\em Nat Microbiol}, 4(11):1885--1894, November 2019.

\bibitem{Wang2017-mb}
J~Wang, M~Brackmann, D~Casta{\~n}o-D{\'\i}ez, M~Kudryashev, K~N Goldie,
  T~Maier, H~Stahlberg, and M~Basler.
\newblock {Cryo-EM} structure of the extended type {VI} secretion system
  sheath-tube complex.
\newblock {\em Nat Microbiol}, 2(11):1507--1512, November 2017.

\bibitem{Guerrero-Ferreira2019-on}
R~C Guerrero-Ferreira, M~Hupfeld, S~Nazarov, N~M Taylor, M~M Shneider, J~M
  Obbineni, M~J Loessner, T~Ishikawa, J~Klumpp, and P~G Leiman.
\newblock Structure and transformation of bacteriophage {A511} baseplate and
  tail upon infection of listeria cells.
\newblock {\em EMBO J.}, 38(3), February 2019.

\bibitem{Coombs1994-xa}
D~H Coombs and F~Arisaka.
\newblock {T4} tail structure and function.
\newblock {\em Molecular biology of bacteriophage}, 1994.

\bibitem{Aksyuk2009-wp}
A~A Aksyuk, P~G Leiman, L~P Kurochkina, M~M Shneider, V~A Kostyuchenko, V~V
  Mesyanzhinov, and M~G Rossmann.
\newblock The tail sheath structure of bacteriophage t4: a molecular machine
  for infecting bacteria.
\newblock {\em EMBO J.}, 28(7):821--829, April 2009.

\bibitem{Taylor2018-fn}
N~M~I Taylor, M~J van Raaij, and P~G Leiman.
\newblock Contractile injection systems of bacteriophages and related systems.
\newblock {\em Mol. Microbiol.}, 108(1):6--15, April 2018.

\bibitem{Arisaka1981-uq}
F~Arisaka, J~Engel, and H~Klump.
\newblock Contraction and dissociation of the bacteriophage {T4} tail sheath
  induced by heat and urea.
\newblock {\em Prog. Clin. Biol. Res.}, 64:365--379, 1981.

\bibitem{Leiman2004-fm}
Petr~G Leiman, Paul~R Chipman, Victor~A Kostyuchenko, Vadim~V Mesyanzhinov, and
  Michael~G Rossmann.
\newblock Three-dimensional rearrangement of proteins in the tail of
  bacteriophage {T4} on infection of its host.
\newblock 118:419--429.

\bibitem{To1969-es}
C~M To, E~Kellenberger, and A~Eisenstark.
\newblock Disassembly of {T}-even bacteriophage into structural parts and
  subunits.
\newblock 46:493--511.

\bibitem{Leiman2012-ps}
P~G Leiman and M~M Shneider.
\newblock Contractile tail machines of bacteriophages.
\newblock {\em Adv. Exp. Med. Biol.}, 726:93--114, 2012.

\bibitem{Fraser2021-zu}
A~Fraser, N~S Prokhorov, F~Jiao, B~M Pettitt, S~Scheuring, and P~G Leiman.
\newblock Quantitative description of a contractile macromolecular machine.
\newblock {\em Sci Adv}, 7(24), June 2021.

\bibitem{Maghsoodi2017-lt}
A~Maghsoodi, A~Chatterjee, I~Andricioaei, and NC~Perkins.
\newblock Dynamic model exposes the energetics and dynamics of the injection
  machinery for bacteriophage {T4}.
\newblock {\em Biophys. J.}, 113(1):195--205, July 2017.

\bibitem{Maghsoodi2019-ym}
A~Maghsoodi, A~Chatterjee, I~Andricioaei, and N~C Perkins.
\newblock How the phage {T4} injection machinery works including energetics,
  forces, and dynamic pathway.
\newblock {\em Proc. Natl. Acad. Sci. U. S. A.}, 116(50):25097--25105, December
  2019.

\bibitem{Ge2020-kx}
P~Ge, D~Scholl, N~S Prokhorov, J~Avaylon, M~M Shneider, C~Browning, S~A Buth,
  M~Plattner, U~Chakraborty, K~Ding, P~G Leiman, J~F Miller, and Z~H Zhou.
\newblock Action of a minimal contractile bactericidal nanomachine.
\newblock {\em Nature}, 580(7805):658--662, April 2020.

\bibitem{Trefethen2000-wg}
L~N Trefethen.
\newblock {\em Spectral methods in {MATLAB}}.
\newblock Software, environments, tools. Society for Industrial and Applied
  Mathematics, Philadelphia, Pa., 2000.

\bibitem{burns2020}
Keaton~J Burns, Geoffrey~M Vasil, Jeffrey~S Oishi, Daniel Lecoanet, and
  Benjamin~P Brown.
\newblock Dedalus: A flexible framework for numerical simulations with spectral
  methods.
\newblock {\em Physical Review Research}, 2(2):023068, 2020.

\end{thebibliography}
\bibliographystyle{unsrt} 

\end{document}